\documentclass[AMA,Times2COL]{WileyNJDv5} 

\makeatletter
\gdef\@dummy@received{}%
\gdef\@dummy@revised{}%
\gdef\@dummy@accepted{}%
\gdef\@dummy@pubdate{}%
\gdef\@received{}%
\gdef\@revised{}%
\gdef\@accepted{}%
\gdef\@published{}%
\gdef\@history@dates{}%
\gdef\@DOI@text{}       
\renewcommand{\doiheadtext}[1]{} 
\def\oddfoot@titlepage@info{}%
\def\evenfoot@titlepage@info{}%
\makeatother

\articletype{Research Article}%

\received{Date Month Year}
\revised{Date Month Year}
\accepted{Date Month Year}
\journal{Journal}
\volume{00}
\copyyear{2023}
\startpage{1}

\begin{document}

\title{Evaluating Idle Animation Believability: a User Perspective}

\author[1]{Eneko Atxa Landa}

\author[1]{Elena Lazkano}

\author[1]{Igor Rodriguez}

\author[1]{Itsaso Rodriguez-Moreno}

\author[1]{Itziar Irigoien}

\authormark{ATXA \textsc{et al.}}
\titlemark{Evaluating Idle Animation Believability: a User Perspective}

\address[1]{\orgdiv{Computational Science and Artificial Intelligence}, \orgname{University of the Basque Country}, \orgaddress{\state{Gipuzkoa}, \country{Spain}}}

\corres{Corresponding author: Eneko Atxa Landa. \email{eneko.atxa@ehu.eus}}

\presentaddress{Computational Science and Artificial Intelligence, University of the Basque Country, Faculty of Computer Science, Donostia/ San Sebastián}


\abstract[Abstract]{Animating realistic avatars requires using high quality animations for every possible state the avatar can be in. This includes actions like walking or running, but also subtle movements that convey emotions and personality. Idle animations, such as standing, breathing or looking around, are crucial for realism and believability. In games and virtual applications, these are often handcrafted or recorded with actors, but this is costly. Furthermore, recording realistic idle animations can be very complex, because the actor must not know they are being recorded in order to make genuine movements. For this reasons idle animation datasets are not widely available. Nevertheless, this paper concludes that both acted and genuine idle animations are perceived as real, and that users are not able to distinguish between them. It also states that handmade and recorded idle animations are perceived differently. These two conclusions mean that recording idle animations should be easier than it is thought to be, meaning that actors can be specifically told to act the movements, significantly simplifying the recording process. These conclusions should help future efforts to record idle animation datasets. Finally, we also publish ReActIdle, a 3 dimensional idle animation dataset containing both real and acted idle motions.
}

\keywords{Motion Capture, Idle Motion, Motion Perception, Animation}

\jnlcitation{\cname{%
\author{Eneko Atxa Landa},
\author{Elena Lazkano},
\author{Igor Rodriguez},
\author{Itsaso Rodríguez-Moreno}, and
\author{Itziar Irigoien}}.
\ctitle{Evaluating Idle Animation Believability: a User Perspective} \cjournal{\it Computer Animations and Virtual Worlds} \cvol{2021;00(00):1--18}.}

\maketitle

\renewcommand\thefootnote{}

\renewcommand\thefootnote{\fnsymbol{footnote}}
\setcounter{footnote}{1}

\section{Introduction}
Realism in virtual agents is affected by several aspects such as the quality of the 3D models used, the human-like behaviour of the avatar or having lifelike animations. It is also certainly an easy illusion to break: a perfectly modelled character, with a truly human-like behaviour and a realistic way of speaking can suddenly become obviously artificial if, for instance, it suddenly freezes completely after it has finished speaking. To maintain the illusion of realism, the character must have realistic behaviour at all times, with no stops or cuts in between actions.

That is where idle animations come into play. These form a group of animations depicting subtle, sometimes imperceptible actions, such as breathing, looking around, making small body movements or scratching body parts. They maintain the realism of the agent, even if nothing is directly interacting with them. Such is their significance, that highly realistic virtual worlds and virtual interactive environments, especially games, implement handcrafted or pre-recorded idle animations for every character in them. However, creating these animations increases the cost of creating virtual experiences, which can pose a significant barrier for independent developers and small studios with limited budgets.

In the field of motion generation, research areas such as co-speech gesture generation, 3D human motion prediction, and human motion generation are actively supported by readily available high-quality datasets. These datasets facilitate advances within these research domains. However, the availability of public good quality idle animation datasets is very scarce. One possible reason for this scarcity is the perceived complexity of capturing genuine idle movements. Firstly, recording truly natural idle behaviour requires capturing subjects ``in the wild'', meaning they must be unaware of the fact that they are being observed. This is crucial as conscious awareness of being recorded would change the way they move, resulting in ungenuine movements. The recording process also presents a significant challenge: on the one hand, obtaining informed consent from individuals after the recordings is ethically problematic, and on the  the other hand, once a subject has been recorded without their prior knowledge, they cannot be recorded again using the same method due to the potential psychological conditioning in the way they behave. Additionally, the use of motion capture suits is incompatible with capturing genuine idle movements. These suits generally permit recording high quality data with little noise, but their use is intrusive, because they alert the subject of the recording process and thus have an effect on the genuineness of the animations.

In an effort to mitigate the insufficiency of idle animation datasets, we carried out an analysis to disprove the hypothesis that idle animation has to be recorded in a non-intrusive way. In this work, we investigate the perception of idle animations in virtual characters through user studies and variable analysis. On the one hand, we compare acted and genuine idle movements, and on the other hand, handmade and recorded movements.

The process to analyse the perception of idle animations has been the following: firstly, we recorded a dataset in 3 dimensions, containing both genuine and acted idle motion. Secondly, we designed a user study in which users were shown videos containing 3D renders of genuine and acted idle animations in random order, and they were asked to classify each video in one of these two groups. The analysis of the answers has shown that users cannot correctly discriminate between genuine and acted idle animations. We have also analysed the motion data directly to compare the data distributions in terms of joint and angular speeds. The analysis has also been extended to examine the perception of handmade idle animations in comparison to the animations from the recorded dataset.

\section{Related work} \label{sec:relatedWork}
Automatic gesture generation is a vast field of research, encompassing different tasks. In this section, an overview of some automatic gesture generation tasks and their corresponding datasets is presented. These research fields incorporate many different modalities in motion generation, such as co-speech gestures, motion prediction or text conditioned generation. The variety, quantity and quality of datasets that they have has lead to the development of generative systems in each field. Finally, some works and datasets of the idle gesture generation field are presented.

\subsection{Co-speech gesture generation}
There are multiple papers regarding co-speech gesture generation. Qi et al. \cite{qi2023emotiongesture} propose \textit{EmotionGesture}, a framework that generates 3 dimensional gestures from audio and consists of an Emotion-Beat Mining module and a Spatial Temporal Prompter module to solve the task. In another instance, Yi et al. \cite{yi2023generating} present \textit{TALKShow}, generating both body and hand animations as well as face animations over a 3 dimensional mesh. In \textit{DiffGesture}, \cite{zhu2023taming} diffusion models are proposed to effectively capture the cross-modal audio-to-gesture association and preserve temporal coherence. Yearly, new methods for generation are presented on the Genea Challenge \cite{kucherenko2023genea}. We refer the reader to \cite{nyatsanga2023comprehensive} for a more comprehensive survey on co-speech gesture generation.

The available datasets in this field vary in size, dimensionality, number of modalities, or whether they are monologues or dialogues, for example. Among some of the most recent datasets, \textit{BEAT} \cite{liu2022beat} contains 76 hours of high-quality multimodal data of 30 speakers talking with eight different emotions and in four different languages. It also has frame-level annotations on emotion and semantic relevance, containing 32 million annotations that complement the motion data with other modalities that could be interesting regarding gesture generation. Other co-speech gesture generation datasets are \textit{Talking with hands} \cite{lee2019talking} or \textit{ZEGGS} \cite{ghorbani2023zeroeggs}. The aforementioned review \cite{nyatsanga2023comprehensive} lists and analyses the available datasets and their typology.

\subsection{Human motion prediction}
Human motion prediction is also a crucial task regarding automatic gesture generation, which consists in continuing sequences of different human movements or actions. Martinez et al. \cite{martinez2017human} train a simple RNN to generate different types of motion based on another starting sequence. Newer architectures have also been proposed: in \cite{cui2021efficient}, Cui et al. present a Temporal Convolutional GAN to forecast future poses, and model long-term dependencies by using hierarchical temporal convolution. Lyu et al. \cite{lyu2021learning} model the motion prediction problem with stochastic differential equations and path integrals simulated by using GANs. A more comprehensive review of the human motion prediction task by Lyu et al. can be found in \cite{lyu20223d}.

The review also analyses many datasets for the motion prediction task: for instance, \textit{Human3.6M} \cite{ionescu2013human3} contains 3.6 million frames of 3 dimensional human poses performing 17 different tasks, such as smoking, taking photos or talking on the phone. The \textit{CMU Panoptic} dataset \cite{hanbyuljoo2019panoptic} is a 3 dimensional dataset containing 1.5 million 3D poses in 5.5 hours of data. It contains scenes with one or many people in them, interacting and carrying out different actions, such as dancing, playing instruments, interacting in social activities, playing or cleaning a room. Additionally, \textit{AMASS} \cite{mahmood2019amass} unifies 15 different optical marker-based datasets by representing them within a common framework and parametrisation, containing 40 hours of motion data, more than 300 subjects and more than 11,000 motions.

\subsection{Text conditioned human motion generation}
Text conditioned human motion generation is a research field that aims to generate human motion sequences from textual descriptions. It is a complex task that requires both understanding natural language and human motion and the relations between them, by converting the descriptions to animations. Zhu et al. describe the text conditioned human motion generation field in a more general human motion generation survey \cite{zhu2023human}, alongside other motion generation tasks, such as the aforementioned co-speech gesture generation or music to dance generation. Some recent motion generative models include \textit{Motion Mamba} \cite{zhang2025motion}, which uses state space models (SSMs) to model long sequences of motion efficiently. \textit{MotionDiffuse} \cite{zhang2022motiondiffuse} is a diffusion model which generates text-driven motion, that has probabilistic language-motion mapping, realistic motion synthesis and multi-level manipulation.

Text conditioned generation also has many available datasets. \textit{BABEL} \cite{punnakkal2021babel} contains 43 hours of motion capture sequences from \textit{AMASS}, alongside action labels. \textit{HumanML3D} \cite{guo2022generating} is another dataset that combines the \textit{HumanAct12} and the \textit{AMASS} datasets into a nearly 29 hour dataset, and provides text descriptions for the sequences. As can be seen, many datasets for this field derive from others used in human motion prediction, by expanding them with useful labels. The aforementioned human motion generation survey by Zhu et al. describes the most important datasets in the field.

\subsection{Idle motion generation} \label{subsec:idle_motion}
However, when it comes to idle motion, the quantity of scientific literature drastically decreases. Egges et al. \cite{egges2004personalised} created an idle motion engine based on Principal Component Analysis, which generates motion by combining small posture variations and change of balance. Egges et al. \cite{egges2004example} further developed an idle motion engine with a GUI, centred in blending pre-recorded animations, but again, centred on small posture variations and balance change, which can be restrictive. Kocoń \cite{kocon2020head} developed an idle motion synthesiser on a 3D human head model, by using kinematic chains of rigid elements which generate idle movements.

The impact of idle animation on social robotics is also notable, since there are many works which apply them in social robots, measure their impact, and finally emphasise their importance. Cuijpers et al. \cite{cuijpers2015motions} analyse idle and meaningful motions in robots through the lens of social verification. Song et al. \cite{song2009design} manually design idle movements for a specific service robot and measure how people perceive and interpret them. Asselborn et al. \cite{asselborn2017keep} explore the effect of idle motions in anthropomorphism and robot engagement on children.

In the idle animation generation task, there are nearly no public datasets to work with. \textit{IdlePose} \cite{ravenet2021idlepose} is the only reported idle data collection effort. They recorded idle motion in the most spontaneous way possible by tricking people into thinking they were recording another type of dataset, therefore recording idle motion unknowingly. Even if the used strategy is very clever and the ethical issues are brilliantly addressed, the downside is that it only contains 2 dimensional data, as it uses a single conventional camera and pose estimation software that works in 2 dimensions. That does not permit using the data for 3D animation, which is nowadays the one that is used in most virtual applications. 

Mixamo \cite{mixamoMixamo} is another open library that contains many animations and 3D models that can be used for many computer applications such as making games. This source does contain some handmade idle animations in 3 dimensions, but most of them are very specific (for example, idle animations holding a gun or crouching) and not useful for general applications. They are also short, ranging from 100 to 250 frames, as they are prepared to be easily looped and transitioned to other animations. Even if the dataset returns more than 500 results for the word ``idle'', after filtering out specific idle animations, transitions and repeated animations, 15 animations were considered useful for general situations. Finally, the content in Mixamo can be used for research, it cannot be used to train machine learning models according to their terms of use, which can be a limitation for specific research projects.

Therefore, the need for a large 3D dataset containing idle animations is crucial in order to be able to develop research in this area. Analysing whether it is essential to capture genuine ``in the wild'' idle motion or not will help by providing guidelines for future dataset recording efforts. Simplifying the recording procedures could result in more research being carried out in idle motion generation.

\section{Data recording and processing} \label{sec:dataRecording}
When recording the data, since we wanted to disprove the hypothesis presented above (i.e. that idle animations should be captured in a genuine manner and not by acting them), it was very important to establish a non-intrusive recording procedure, provided that we needed  to collect both real and acted idle motion. In order to record genuine idle motion in the least intrusive way possible, the recording equipment was also selected with this in mind: it consists of several conventional cameras and an open source software called Freemocap \cite{matthis2022freemocap}. In this section, we first describe the hardware and software setup used, and then the recording procedure.

\subsection{Hardware and software setup}
Motion capture suits are extremely intrusive as the person being recorded is obviously aware of wearing one and they do not enable the subject to make natural movements. For this reason, we used a multiple viewpoint and triangulation based setting: we used a 4 camera setup, combined with the Freemocap software which detects 2D skeletons from multiple synchronised videos and later uses triangulation to combine these skeletons into a 3D representation. This enables recording 3D animations without the need of expensive depth cameras or depth estimation models.

To record the individual videos, we used 4 Logitech c920 webcams located in a semicircle around a marked area in which the actors would be placed. The cameras recorded four simultaneous videos at $720 \times 1280$ resolution and 30 frames per second (fps). The videos need to be synchronised for the software to work properly, so this was later manually executed with the help of a clapperboard.

\begin{figure}[!htbp]
  \centering
  \includegraphics[width=\linewidth]{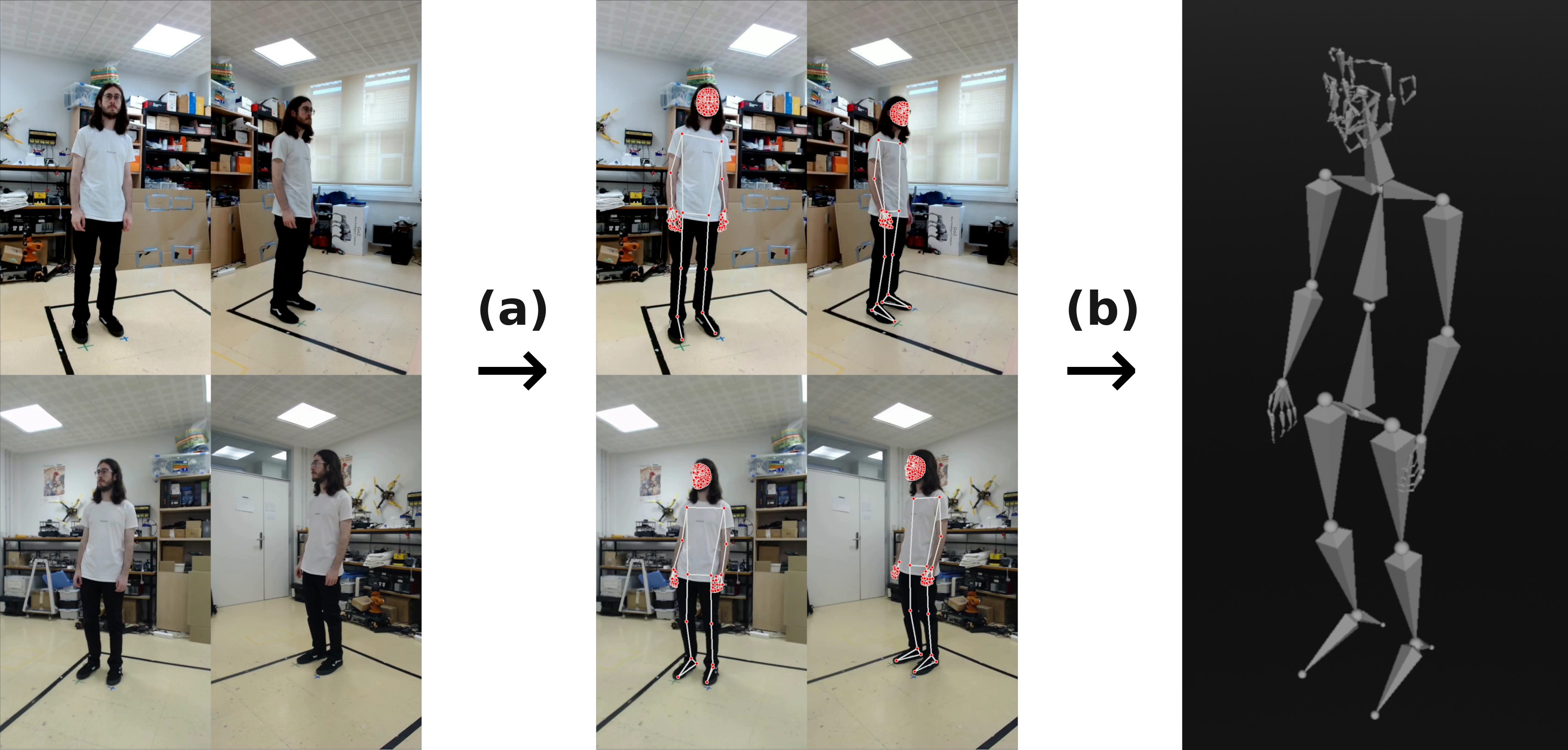}
  \caption{\label{fig:freemocap}Freemocap first detects the 2D skeletons using Mediapipe (a) and then uses triangulation (b) to create a 3D skeleton.}
\end{figure}

Afterwards, the videos were processed with Freemocap to extract 3 dimensional motion capture data. The 4 webcams were calibrated with a ChArUco board in order to parametrise a 3 dimensional space. This geometric information is stored in a configuration file for later use in the inference. Then, as shown in Figure \ref{fig:freemocap}, the software uses the Mediapipe \cite{lugaresi2019mediapipe} pose estimation pipeline to detect 2 dimensional skeletons in each video, and lastly uses triangulation to recreate the final 3 dimensional skeleton. The final animations were exported in the Biovision Hierarchy (BVH) format, which is a very widely used format in animation. We believe this is currently one of the best ways to record motion capture with the minimal amount of intrusion possible.

\subsection{Recording procedure}
The recorded data consists of 2 types of different motion: genuine idle motion and acted idle motion.

\textbf{Genuine idle motion:} this part, i.e. the genuine idle motion recording, is the most crucial one. Ideally, it should be conducted without the actors knowing that they are being recorded, so they make the most genuine movements possible. A similar deception process to that in \cite{ravenet2021idlepose} was carried out, which has ethical considerations that have been considered and resolved, which are detailed below.

The person was brought to the recording area and tricked into thinking that before starting the recording session, a small synchronisation process had to be performed with the audio and the video. They were told to wait in silence, since the audio synchronisation needed silence in order to work. The subject was unknowingly being recorded for 2 minutes, resulting in genuine idle movements.

To address the ethical implications of deceiving a person and recording them without them knowing, after the whole recording process was finished, every participant was given an explanation that the first synchronisation process was, in fact, false, and the real purpose of that part was revealed to them. The objective of the recording was made clear to them, and the need for the first part to be secret was explained. The option to withdraw the recording was presented, although no participant decided to withdraw any of them. The experiment was evaluated and accepted by the necessary ethical committee.

However, the procedure did not work with every participant. Since to get the most genuine motion possible, they were not restricted by any rules, some participants may have had taken their phone out, moved from the recording area or spoken out loud. Those recordings have been discarded from the final data.

\textbf{Acted idle motion:} in this part, the participants were instructed to act as if they were waiting for someone or waiting for a bus on the street. Minimal intervention is optimal for this part, but they were told not to use their phones or move away from the recording area, in order to have clean idle motion. This part also lasted for another 2 minutes.

\textbf{Final dataset specifications:} finally, a manual processing phase was conducted. Noisy and faulty data was removed, and many entire recordings from the genuine idle motion were discarded. The final dataset contains 27,273 frames or 15.15 minutes of genuine idle motion and 55,039 frames or 30.57 minutes of acted idle motion. The dataset containing both the acted and the genuine idle animations alongside the code to reproduce the results from this paper will publicly available upon acceptance. Some sample animations are sent as supplementary material to revise the quality of the animations.

\section{Comparative analysis of genuine and acted idle motion}\label{sec:analysis1}
After recording the  genuine and acted idle motion, a thorough analysis was carried out to compare the data in the two distributions. The main part of the analysis is a user study conducted to analyse whether humans are able to differentiate between these two types of idle motion or they are perceptually the same. We also conducted a direct analysis over the actual positional and rotational data to compare the two distributions in terms of average joint and angular speeds.

\subsection{User study}
\begin{figure}[!htbp]
  \centering
  \includegraphics[width=\linewidth]{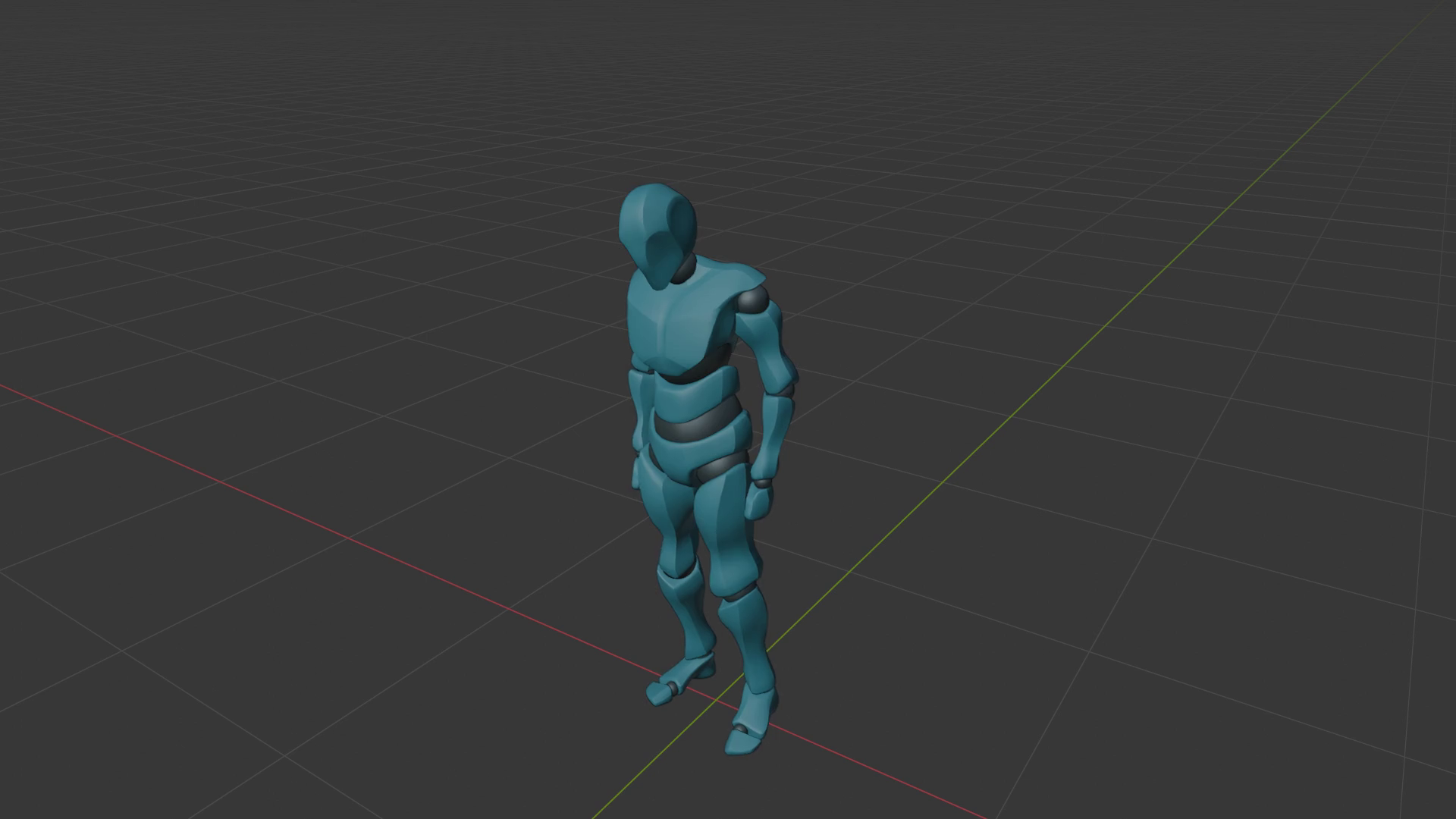}
  \caption{\label{fig:exampleFrame}The videos showed a 3D render of the animations using a neutral model downloaded from Mixamo.}
\end{figure}

We designed a user study that consisted on showing renders of the two types of recorded data to the participants and measuring whether they were able to distinguish between the two classes of motion. The study was conducted on 123 participants. 30 videos of 10 seconds each were shown to all of them in random order (15 real and 15 acted animations). In each video, there was a real or an acted idle animation piece, randomly selected from the dataset. They were explained how the data was recorded in the two situations, and for each video, the participant had to classify the animation in the video as ``real'' or ``acted''. The videos contained 3D renders of the recorded BVH files, applied to a 3D model of a humanoid. The model was the ``Y bot'' model downloaded from Mixamo, selected to be the most neutral model possible. The fingers of the model were removed since the motion capture did not provide high quality finger data, and could distract or condition the users. The reason for the videos being rendered using the same geometric model is that the geometric model impacts the perception in humans \cite{hodgins1998perception}. All the other possible rendering variables are also kept the same for all the videos, such as model size, camera view or lighting conditions. Figure \ref{fig:exampleFrame} shows a frame from one of the videos that were shown. It is important to emphasise that the participants were not told anything about how to differentiate between the two labels: we wanted to measure the inherent capability of people to perceive and classify the animations. 
In addition to the task in hand, all users were asked for some information to measure their level of technology usage: they were enquired about which social networks they used and whether or not they played or used to play video games. After the task ended, they rated the perceived difficulty of the task in a Likert scale form 1 (very easy) to 5 (very difficult).
\subsubsection{Results} \label{subsub:results1}
\begin{figure}[!htbp]
  \centering
  \includegraphics[width=\linewidth]{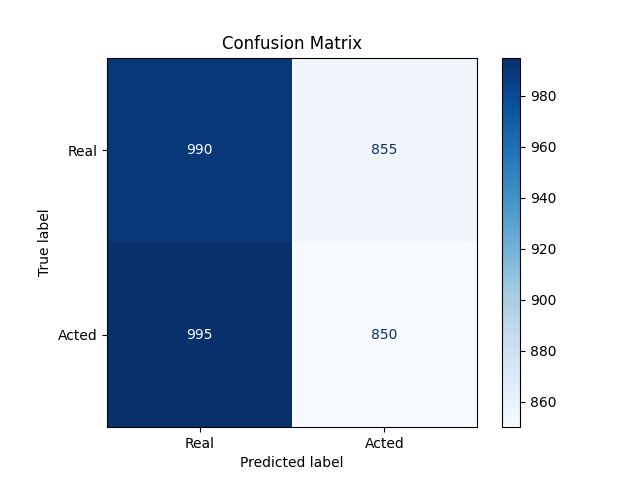}
  \caption{\label{fig:real_acted_confusion}The results from the first user study show very similar answer proportions for real and acted animations. This suggests that the participants are not able to correctly differentiate real and acted instances.}
\end{figure}

The confusion matrix created from the obtained results (Figure \ref{fig:real_acted_confusion}) shows the counts of predicted and real labels of the videos and does not show any remarkable pattern. Among real animations, the proportions of the answers were 0.537 and 0.463 for real and acted, respectively. For acted animations, the proportions were 0.539 and 0.461 for real and acted labels, respectively. This means that users answered similarly for real and acted idle animations. The Chi Square Independence test confirms the results ($p$-value $0.8949$).

Finally, in order to determine whether there is any association between the success rate and the characteristics of the users, we calculated the corresponding correlations and t-tests. We found that there is no strong association between the success rate and any characteristic. The measured attributes were gender, the number of social networks used, whether they played or used to play video games or not, the perceived difficulty and the time needed to finish the test. It is also noteworthy that the average perceived difficulty was $3.88$.

\subsection{Analysis of the motion variables}
We also analysed the general variables of the two types of idle motion, to see if there is any divergence in the internal values that define the motion. For each skeleton joint, we compared the average joint velocities of the two distributions. We also compared the average rotational velocities of the two distributions, for each joint rotation. Directly analysing the rotational variables permits to ignore the skeleton size, to just attend to the rotational values.

\begin{figure}[!htbp]
  \centering
  \includegraphics[width=\linewidth]{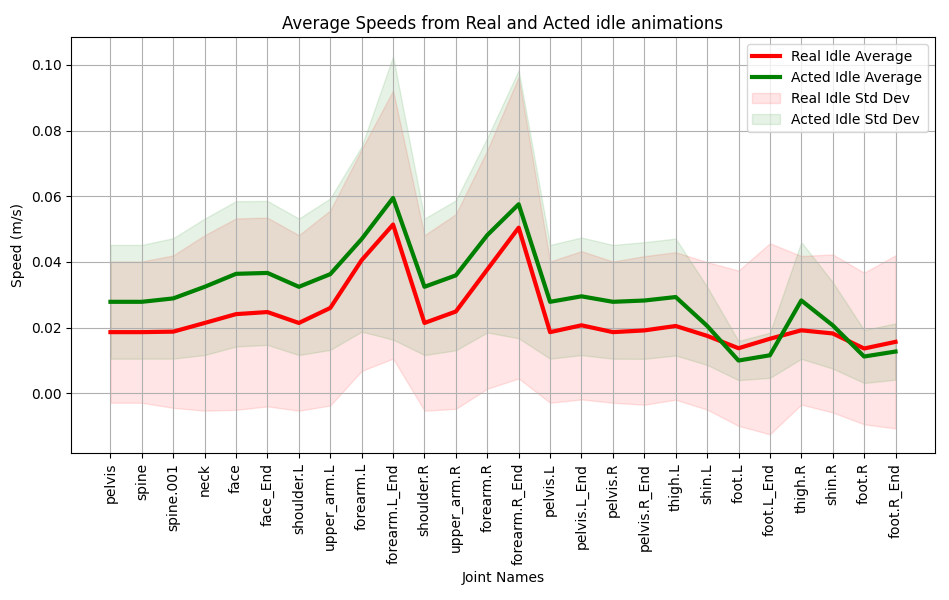}
  \caption{\label{fig:speeds}Average speed for each joint, divided in real and acted animations.}
\end{figure}

\begin{figure}[!htbp]
  \centering
  \includegraphics[width=\linewidth]{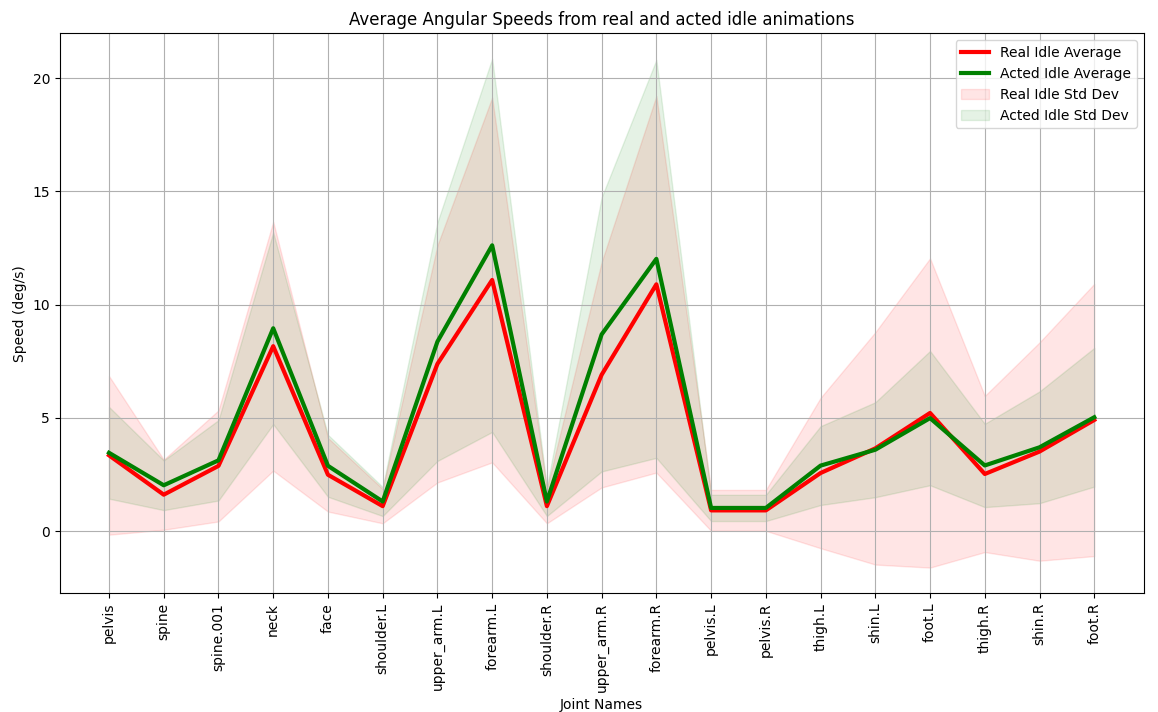}
  \caption{\label{fig:angular_speeds}Average angular speed for each joint, divided in real and acted animations.}
\end{figure}

Figure \ref{fig:speeds} shows the average speed for each joint, divided in real and acted idle motions. Both types of motions have extremely similar average speeds for all joints. Generally, it can be seen that the arms are the limbs that have the highest speeds in idle animations, especially in the hands, as expected. The standard deviation is smaller in the feet in acted animations, but the expected value is still very similar. The same pattern can be seen for angular speeds in figure \ref{fig:angular_speeds}, although the neck also seems to have higher angular speeds than other limbs. This is reasonable, as a high angular movement in the neck does not have a big impact on the head joint position, as the limb in itself is shorter than other limbs, such as arms. The leg joints show a narrower standard deviation, but still a very similar average. In conclusion, there is no relevant difference between genuine and acted idle animations in joint speed or angular speed for any joint.

\section{Comparative analysis of recorded and handmade idle motion}\label{sec:analysis2}
In addition to analysing the perception of real and acted idle motion, we also analysed the perception of recorded and handmade idle animations to determine whether the method of creating the animations influences how users perceive the credibility of these motions. Handmade animations are a type of animation widely used in video games, as they can be specifically modified according to the needs of each application, and they are very controllable in a fine-grained level. In this way they enable to create easy transitions between different handmade animations, because many aspects of the handmade animations are known, such as joint speeds in each frame or animation cycle lengths.

However, we wanted to test whether they are perceptually similar to motion capture data, as the latter may have a different level of realism because it comes from real actors. It is noteworthy that a perceptual difference does not mean that one animation type is better than the other, it only means that people are able to distinguish between the two types, so this would need to be taken into account if the two idle animation types were to be mixed. 
\subsection{User study}
We designed a second user study that consisted on showing participants renders of handmade animations and renders of recorded idle sequences, measuring again whether they were able to distinguish between the two classes of motion. For the recorded data, we used the acted portion of the instances in the first experiment. For handmade data, we used Mixamo as the source of the animations. As mentioned in Section \ref{subsec:idle_motion}, we searched for the keyword ``idle'' inside Mixamo, and manually filtered unusable animations (such as transitions, specific situations or repeated animations) to get the largest possible subset of usable general idle animations. Eventually, 15 idle animations were selected for this purpose.

The second user study was carried out on 114 participants. Some of the users participated in both studies, in random order, but the tests were done one week apart from each other, so there was a sufficient time difference between them. The test was exactly the same as in the first experiment, but in this case 15 renders of recorded motions and 15 handmade animations were shown to each participant. They were explained that there were 2 types of motion, and each participant had to classify each video as ``recorded'' or ``handmade''. The render parameters were exactly the same as in the first experiment: we used the same 3D model, lighting, camera view and clip duration. Some video clips from Mixamo had to be looped twice to reach 10 seconds, but this is still a fair comparison because being short and loopable is a typical characteristic of handmade animations. The users were also asked the same questions about themselves: which social networks they used and if they played or used to play video games. After the task ended, they were also asked to rate the perceived difficulty of the task in a Likert scale form 1 (very easy) to 5 (very difficult).
\subsubsection{Results} \label{subsub:results2}
\begin{figure}[!htbp]
  \centering
  \includegraphics[width=\linewidth]{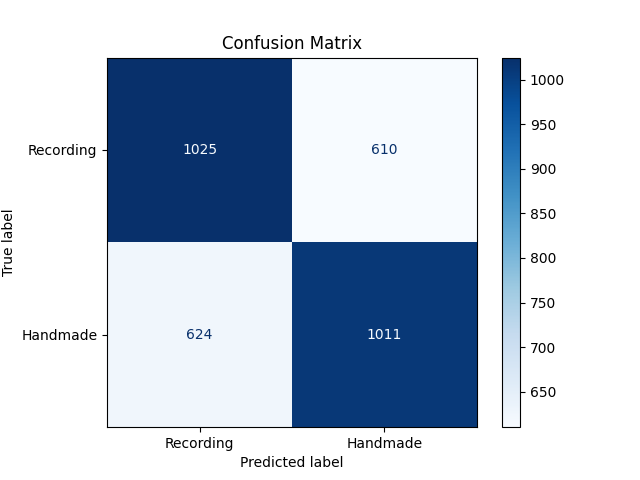}
  \caption{\label{fig:recorded_handmade_confusion}The results from the second user study show different answer proportions for handmade and recorded instances. This suggests that the users were generally able to differentiate between handmade and recorded instances.}
\end{figure}

The confusion matrix created from the obtained results (Figure \ref{fig:recorded_handmade_confusion}) shows the counts of real and predicted labels of the second user study. In this case, the pattern is quite remarkable: the confusion matrix has more responses in the diagonal, meaning that a bigger part of the classification has been done correctly. Among the recorded videos, the proportions of the answers were 0.627 and 0.373 for recorded and handmade animations, respectively. Among handmade videos, the proportions were 0.382 and 0.618 for recorded and handmade animations, respectively. Thus, in general terms, users were able to discriminate between handmade and recorded idle animations. The Chi Square Independence test confirmed the results ($p$-value $1.7825 \cdot 10^{-44}$).

Again, in the second experiment, we observed the association between the success rate and the user's characteristics, and no strong correlations were found between the success rate and any of these. On another note, the average perceived difficulty of the task was $4.02$, higher than in the first one, even if the results were better in the second experiment.
\subsection{Analysis of the motion variables}
\begin{figure}[!htbp]
  \centering
  \includegraphics[width=\linewidth]{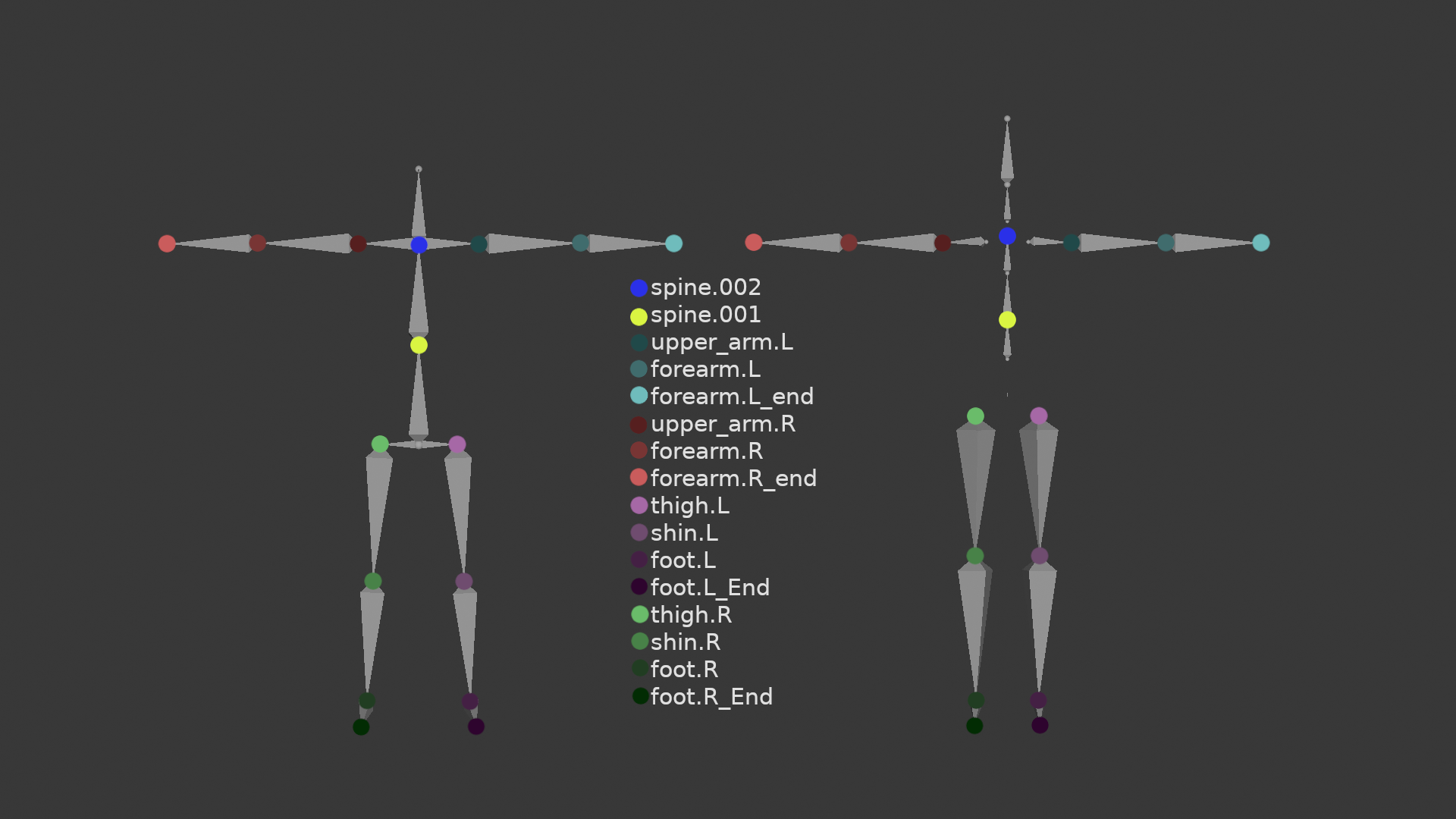}
  \caption{\label{fig:mixamo_freemocap_points_colour}The left skeleton comes from Freemocap and the right one from Mixamo. The coloured points are the points whose speeds can be directly compared. The grey points have no direct translation from one skeleton to the other, so they have to be discarded.}
\end{figure}

In this case, comparing the main variables from recorded and handcrafted motion is not as straightforward as in the first experiment. The available animations from Mixamo and the recorded animations using Freemocap have a different skeletal structure (see Figure \ref{fig:mixamo_freemocap_points_colour}). This means that the joint positions and bone rotations cannot be compared directly.

Firstly, to compare the joint velocities, the skeleton sizes have to be normalised since they might have different sizes, meaning that the scale of the changes in joint positions could differ. Both of the skeletons have been normalised using their respective height. The top bone in the rig from Mixamo has been discarded because it is just an extension of the head bone and it makes the heights different. Even if this normalisation is not exactly perfect, taking into account that the two skeletons have different structures, it is a good approximation that enables comparing the velocities.

After the normalisation, we were able to compare the velocities of the joints. However, as the bone skeleton structures are different, some joints have to be discarded. Figure \ref{fig:mixamo_freemocap_points_colour} shows the two skeletons back to back. The points that are coloured can be directly compared given that they appear in the same or very similar positions in both skeletons. At the same time, the joints that are not coloured have been discarded because direct comparison is not possible.

\begin{figure}[!htbp]
  \centering
  \includegraphics[width=\linewidth]{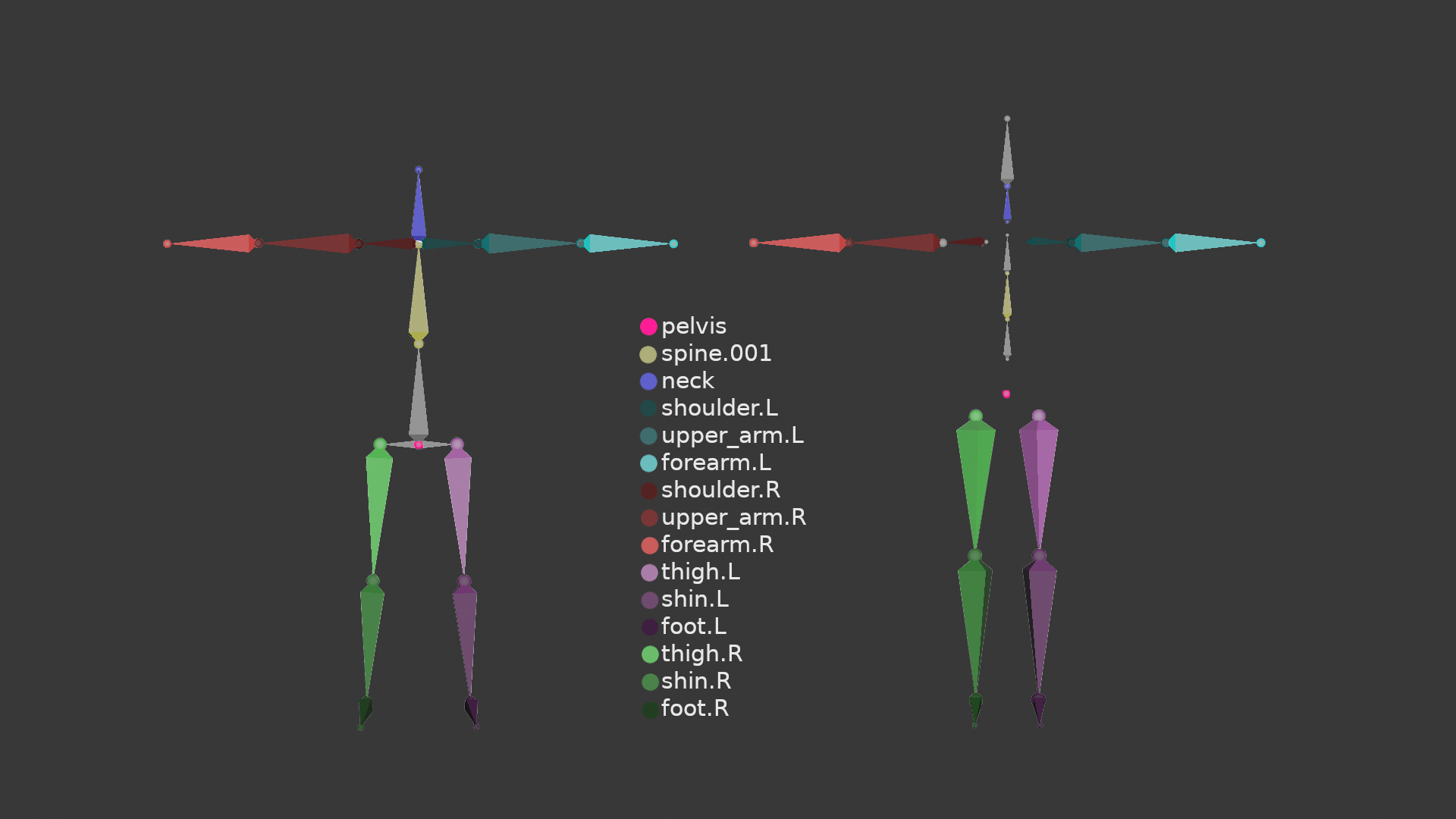}
  \caption{\label{fig:mixamo_freemocap_colour}The skeleton on the left comes from Freemocap and the other one from Mixamo. The coloured bones are the bones whose angular speeds can be directly compared. The grey bones have no direct translation from one skeleton to the other, so they must be discarded.}
\end{figure}

Secondly, to be able to compare the rotational velocities, the scale of the skeleton does not matter, but again, the difference in the bone structures means that not all bones are directly comparable. Figure \ref{fig:mixamo_freemocap_colour} shows the bones whose rotational velocities can be directly compared: each pair of comparable bones is shown in the same colour. The bones that are not coloured have been discarded from the skeletons, as they cannot be easily translated from one skeletal structure to the other.

In addition, it has to be taken into account that even after taking the biggest subset of comparable bones possible, the resting positions of the skeletons are different, meaning that a direct comparison requires an initial transformation from one position to another. Nevertheless, this can be neglected as both speeds and angular speeds are invariant to the initial position if we only take the magnitudes instead of the vectors.

\begin{figure}[!htbp]
  \centering
  \includegraphics[width=\linewidth]{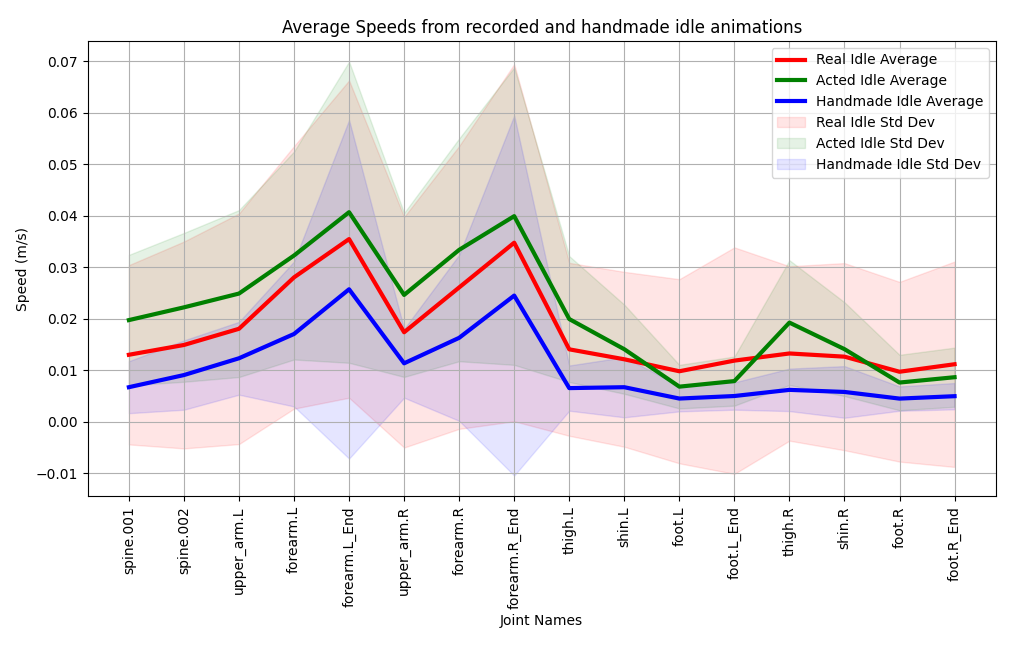}
   \caption{\label{fig:speeds_mixamo}Average velocities of the directly comparable joints. The three curves have a very similar shape, which suggests that joint velocities don't differ between the three motion classes.}
\end{figure}

\begin{figure}[!htbp]
  \centering
  \includegraphics[width=\linewidth]{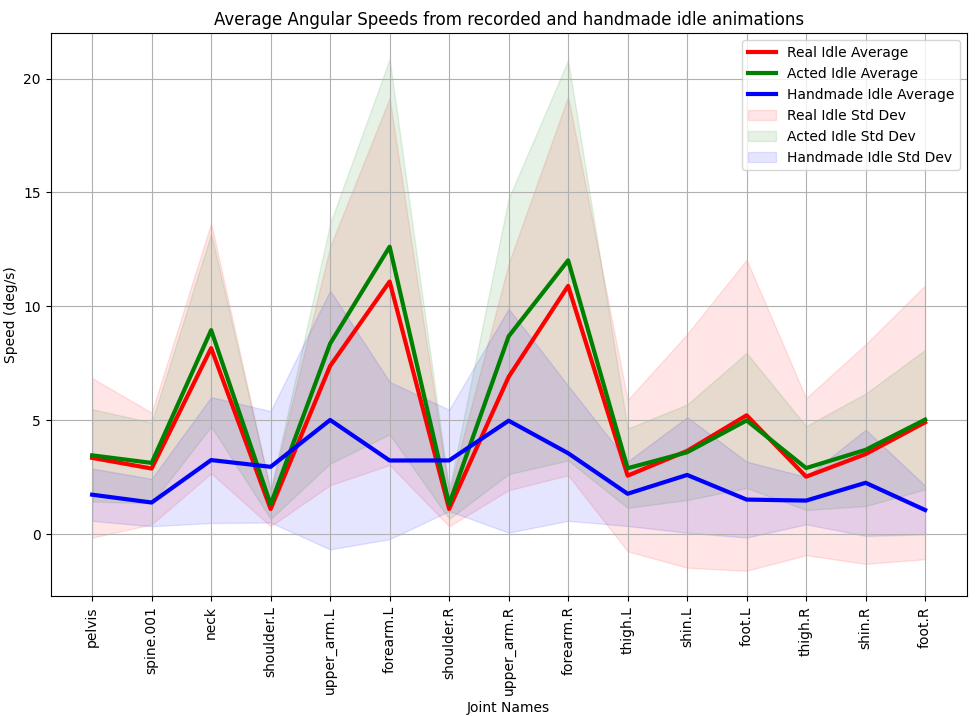}
  \caption{\label{fig:angular_speeds_mixamo}Average angular velocities of the directly comparable joints. The handmade animation curve is more compact overall, and the arms velocities show a different shape. This suggests that forearms move more in recorded data than in the handmade animations.}
\end{figure}

Figure \ref{fig:speeds_mixamo} shows the average speeds for each of the comparable bones. The bones of the Mixamo rig have been renamed to match the ones in the Freemocap rig. Likewise, Figure \ref{fig:angular_speeds_mixamo} shows the average angular speeds.

Firstly, in terms of average speeds of the joints, it can be seen that handmade animations and recorded animations are very similar. Overall, the handmade animations are slower, but the the difference can be considered negligible, as it is similar to the difference between real and acted animations. However, the standard deviation is generally quite smaller in handmade animations, except for the forearms. 

Secondly, Figure \ref{fig:angular_speeds_mixamo} shows that there are some differences in average angular speeds between the handmade and the recorded animations. If we analyse the general tendencies of the data, we can see that on the handmade animations, the angular speeds are more evenly distributed throughout all the joints, and overall, all the average angular speeds are slower, therefore suggesting that the recorded animations contain faster animations. In the same way as in the recordings, the arms have higher average angular speeds than other limbs in handmade animations, but the difference is higher in recorded data. Also, the shape of the angular velocities of the arms show that forearms have higher average angular speeds than upper arms in recorded data, but it is the opposite in the handmade animations. Taking into account that in Figure \ref{fig:speeds_mixamo} the shape of the curves was very similar, this suggests that the arm movement mostly comes from the forearm in the recordings, and the upper arms in the handmade animations. The shape of the angular velocities of the legs differs slightly and the neck has higher angular velocities in the recorded data. However, this could be due to the differences in the structure of the skeleton.

In conclusion, the difference between handmade and recorded animations is more visible than the difference between acted and real motion, but only in terms of angular velocities, and this does not translate to final joint velocities. These differences may occur because of the different skeletal structures and the source of the handmade animations. Generally, we cannot affirm that there is a relevant difference in average speeds, but there is one in average angular speeds, specially in the arms and the neck.
\section{Additional note on accelerations}\label{sec:accelerations}
An analysis on the average acceleration values is not as straightforward as the analysis of average speeds. The average values of accelerations do not easily enable to draw direct conclusions on the analysed motion. However, it does provide an interesting insight into the ``softness'' and the general feeling of the movement of each joint.

\begin{figure}[!htbp]
  \centering
  \includegraphics[width=\linewidth]{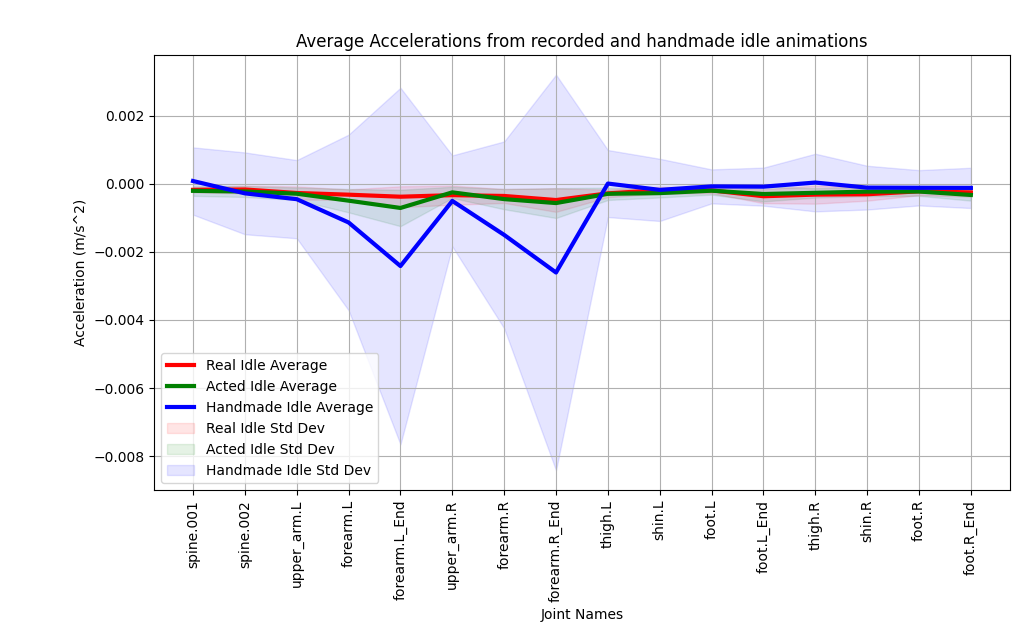}
   \caption{\label{fig:accelerations}Average accelerations of the recorded animations (both real and acted) alongside the average accelerations of the handmade animations from Mixamo.}
\end{figure}

Figure \ref{fig:accelerations} shows the average acceleration values of real, acted and handmade idle motions. In few words, it shows that both for acted and real animations average accelerations are near zero for all comparable joints, whereas for handmade animations, the arms and forearms show a different acceleration pattern. Also, the standard deviation shows very different shapes.

\begin{figure}[!htbp]
  \centering
  \includegraphics[width=\linewidth]{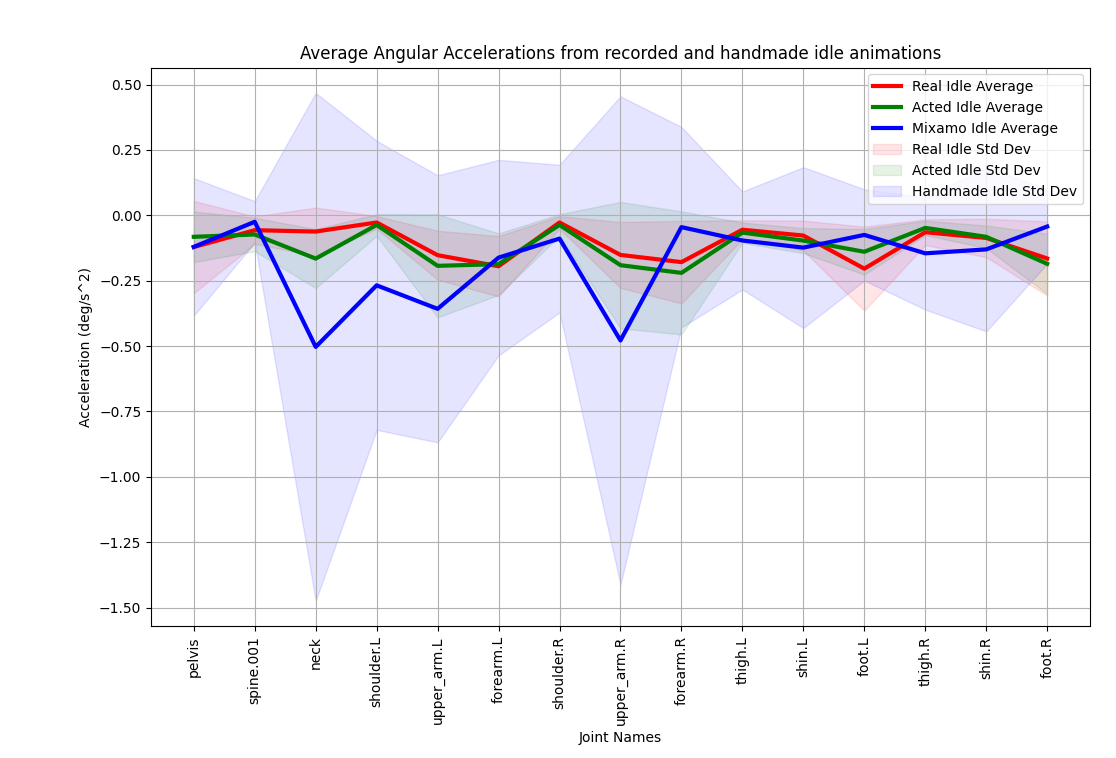}
   \caption{\label{fig:angular_accelerations}Average angular accelerations of the recorded animations (both real and acted) alongside the average angular accelerations of the handmade animations from Mixamo.}
\end{figure}

Figure \ref{fig:angular_accelerations} shows the average angular accelerations for real, acted and handmade motions. Again, for real and acted animations, the pattern is very similar and near zero, but the curve in handmade animations differs. The standard deviation shows a different pattern for handmade animations, too. 

In general, even if direct conclusions cannot be drawn from average acceleration values, it can be seen that real and acted animations are very similar in average acceleration and angular acceleration values, but the pattern differs when comparing to handmade animations, suggesting a difference in the feeling or perception between recorded and handcrafted motions.
\section{Conclusion and future work}\label{sec:conclusions}
In this analysis, we performed an evaluation to measure the perceptive difference between genuine and acted idle motion. In order to do that, we recorded a dataset containing both genuine and acted idle animations, and used the collected data to disprove that idle animations have to be recorded in a genuine manner in order for them to be perceived as real. We performed a user study in which participants had to classify renders from the recorded data between ``real'' and ``acted''. The results suggest that there was no statistically significant difference between the expected and recorded values, therefore implying that users were not able to correctly classify the videos, and thus, the two types of motions are perceptually the same. By directly comparing the average speeds and accelerations of the two motion types, we also showed that there is no relevant difference in the data, either.

We complemented the analysis with another comparison between handcrafted animations from Mixamo and recorded data. Using the same user study methodology, we concluded that these two types of data are, indeed, perceptually different. The users were able to classify the videos correctly, and the difference between the expected and recorded values was statistically significant. The direct analysis of the data supports this conclusion and provides a more detailed insight.

These two findings can be helpful for future attempts of recording idle datasets. Firstly, since acted and real idle animations are perceptually equivalent, we concluded that acted idle animations can be used to create an idle dataset, which simplifies the process of capturing the data. Participants can be asked to act as if they were idling, the situations can be controlled more easily, one actor can be recorded more than once and motion capture suits can be used for precise capture. Secondly, we proved that handmade idle animations are not perceptually the same as recorded animations, so precautions have to be taken when using these two types of motion together.

We also release the recorded 3 dimensional idle animation data. Taking into account that there is currently no public dataset containing long sequences of idle motions in 3 dimensions, we believe that the data could be useful for a variety of applications.

As future work, this analysis can be used as the foundation of an idle dataset recording procedure. We plan on recording a wider dataset of idle animations based on the findings of the study. By simplifying the recording process, recording a larger dataset becomes a more feasible task. We think that having a large idle animation dataset will help in the development of deep-learning based generative models that are able to work with idle animations.

\bmsection*{Acknowledgments}
This work has been partially funded by the Basque Government, Spain, under Research Teams Grant number IT1427-22; the Spanish Ministry of Science (MCIU), the State Research Agency (AEI), the European Regional Development Fund (FEDER), under Grant number PID2021-122402OB-C21 (MCIU/AEI/FEDER, UE); and the University of the Basque Country (UPV/EHU) under grant PIF 23/07.

\bibliography{wileyNJD-AMA}

\appendix

\bmsection{Demographic detail of the user studies\label{app1}}
\vspace*{12pt}
\bmsubsection{User study 1 (Real and acted idle motion)}
Age: Mean = 26.52, Median = 24, SD = 10.97
 
\noindent Gender: Male 68\%, Female 32\%

\noindent Social media usage: YouTube 81.45\%, Instagram 74.19\%, TikTok 27.42\%, Twitter 38.71\%, Facebook 4.84\%

\noindent Video game experience: Yes 66\%, No 34\%

\bmsubsection{User study 2 (Recorded and handmade idle motion)}
Age: Mean = 27.55, Median = 25, SD = 12.57

\noindent Gender: Male 62\%, Female 36\%, Non-binary 2\%

\noindent Social media usage: YouTube 80.7\%, Instagram 70.18\%, TikTok 24.56\%, Twitter 33.33\%, Facebook 7.02\%

\noindent Video game experience: Yes 64\%, No 36\%

\nocite{*}

\end{document}